\documentclass[11pt,onecolumn]{article}
\usepackage{graphicx} 
\usepackage[dvipsnames]{xcolor}
\usepackage[a4paper, total={6in, 8in}]{geometry}

\usepackage{amsmath,amssymb}
\usepackage{xr}
\usepackage{hyperref}
\usepackage[T1]{fontenc}
\usepackage{authblk}

\title{Stray Field NMR: a powerful method to measure dynamics at the millisecond scale}

\author[1,*]{Suzanne LAFON}
\author[2]{Jeanne VEDEL}
\author[2]{Clara TEYNIER}
\author[2]{Divyen RAJ MITHALAL}
\author[1]{Pawel WZIETEK}
\author[1]{Mehdi ZEGHAL}
\author[1,3]{Patrick JUDEINSTEIN}

\affil[1]{Université Paris-Saclay, CNRS, Laboratoire de Physique des Solides, 91405 Orsay, France}
\affil[*]{\texttt{slafon@phare.normalesup.org}}
\affil[2]{Université Paris-Saclay, Ecole Normale Supérieure Paris-Saclay, 91190 Gif-sur-Yvette, France}
\affil[3]{Université Paris-Saclay, CEA, CNRS, Laboratoire Léon Brillouin, 91191 Gif Sur Yvette, France}

\date{}

\begin{document}

\maketitle

\section*{Abstract}

Transport properties in fluids and confined systems play a central role across a wide range of natural and technological contexts, from geology and environmental sciences to biology, energy storage, and membrane-based separation processes. Nuclear Magnetic Resonance (NMR) provides a unique, non-destructive means to probe these properties through species-selective measurements of self-diffusion coefficients. While pulsed field gradient NMR (PFG-NMR) is routinely used, its access to diffusion times is typically limited to values no shorter than about $10$ ms, restricting its applicability to systems with fast dynamics and long relaxation times. Diffusion NMR in a permanent magnetic field gradient (STRAFI) offers a complementary, multiscale approach, enabling diffusion measurements over an extended temporal window, from a few hundred microseconds to several tens of seconds. Despite its strong potential, this technique remains rarely implemented due to experimental and methodological challenges. In this work, we present a robust and versatile STRAFI-based methodology, including a specifically designed experimental setup, optimized pulse sequences, and rigorous data analysis, allowing accurate extraction of self-diffusion coefficients for a broad range of nuclei. The capabilities of the approach are illustrated through diverse applications, including the study of concentrated electrolytes using “NMR-exotic” nuclei ($^{35}$Cl, $^{79}$Br/$^{81}$Br, $^{127}$I, $^{17}$O) and the characterization of micrometre-scale porosity in membranes.



\section{Introduction}

Transport properties of/inside fluids are of utmost importance for natural and technological domains such as geology, pedology, hydrology, environmental sciences, biology and medical sciences, chemical and separation processes in industry, electrochemical devices such as batteries, molecular and macromolecular characterization...  These processes may span on time/space scales ranging from ps/pm to geological scales and may be driven either by thermal agitation or chemical potential or external stimuli, corresponding respectively to self-diffusion or collective motion and advection processes \cite{bird_introductory_2015}.Depending the processes which are involved ant their magnitude, many techniques are already developed and used to characterize these translational  motions \cite{cussler_diffusion_1994}. Beyond them, Nuclear Magnetic Resonance (NMR) techniques based on the control of magnetic field gradient \cite{blumich_2019_essential_nmr} allow to directly measure the diffusion coefficients of individual species with the possibility to measure selectively different nuclei. Most generally these measurements are performed by pulsed field gradient NMR (PFG-NMR) techniques which are now offered as a routine equipment on (nearly) all commercial NMR spectrometers or can be also obtained as accessories when more demanding measurements and conditions are expected \cite{stilbsDiffusionElectrophoreticNMR2019}\cite{price2009nmr}.  These experiments allow also to keep the spectroscopic resolution of NMR spectroscopy in experiments such as DOSY (Diffusion Ordered SpectroscopY), are non destructive, can be performed on samples in controlled environments (temperature, humidity, pressure) and of various geometries and sizes up to tens of centimeters, using (extra) large bore magnet such as those used in magnetic resonance imaging techniques  (MRI) \cite{callaghan2011diffusion}.
They can also be used to measure the behaviour of bulk liquids, dissolved species, or species confined within porous systems \cite{sorland2014diffusion}, providing access to a wealth of information on nanoscale systems \cite{judeinstein_2024_coefficients_diffusion_rmn,judeinstein_2024_nmr_scattering}, since these measurements directly probe the root mean square displacement (RMSD) of spins over the diffusion time, typically corresponding to length scales of $0.1$–$100$ \textmu m. This approach allows determination of both the hydrodynamic radius of molecular/macromolecular species in solution (via the Stokes–Einstein relationship) and the size, geometry, and tortuosity of porous systems. However, due to technical limitations, the minimum accessible diffusion time in PFG-NMR measurements is typically $5$–$10$ ms and cannot be significantly reduced, owing to constraints of the pulse sequence and local perturbations of the homogeneous magnetic field caused by the pulsed gradient. \\

These limitations can be problematic for nuclei with short relaxation times ($T_{1}$ and $T_{2}$), such as confined molecules or quadrupolar nuclei, or when attempting to probe shorter diffusion times in porous or inhomogeneous materials. In such cases, performing NMR experiments in a permanent magnetic field gradient can help to circumvent these issues \cite{stilbsDiffusionElectrophoreticNMR2019}\cite{geilMeasurementTranslationalMolecular1998}. Such experiments may be performed  i) at low field with the permanent gradient of mobile \cite{blumich2020mobile}\cite{Casanova2011} or 1D NMR tomographic magnet setup \cite{judeinsteinLowfieldSinglesidedNMR2017} \cite{panesarMeasurementSelfdiffusionThin2013}, at high field ii) in the natural fringe field (or stray field - STRAFI) domain (STRAFI) of a standard superconducting NMR magnet (out of the central sweet spot) \cite{geilMeasurementTranslationalMolecular1998} \cite{KIMMICH1991136}, or iii) with gradients generated by a Helmholtz coil assembly \cite{privalovTransportMechanismNafion2023}. With these different setups, standard spin echoes sequences are used to measure self-diffusion as it was already highlighted in the NMR pioneering work of Hahn \cite{hahn_spin_1950} and and further developed specifically for the measurement of self-diffusion coefficients \cite{hurlimann_diffusion_2001}\cite{geilMeasurementTranslationalMolecular1998}\cite{KIMMICH1991136}. \\

Although these experiments rely on a relatively simple principle and offer significant potential, their successful implementation requires careful experimental control. The technique is intrinsically characterized by a relatively low signal-to-noise ratio and does not retain the high spectroscopic resolution of DOSY-type NMR experiments. It also calls for specific technical expertise, such as precise control of the probe position and the use of high-quality, frequency-tunable NMR coils. Finally, extracting reliable physical parameters requires high-quality signals, typically obtained through long acquisition times combined with excellent mechanical and electronic stability, to allow reliable data for the fitting process, since the signal decays are described by mathematical relationships involving two or three intertwined variables ($D$, $T_1$ and $T_2$). \\

In this respect, this paper presents a robust and versatile experimental approach based on a dedicated setup, together with a detailed methodology to accurately extract self-diffusion coefficients over a wide range of nuclei through appropriate pulse sequences and data analysis. Several illustrative examples are provided, highlighting the interest of accessing diffusion time scales ranging from a few tens of microseconds to several tens of seconds. This extended temporal window enables, in particular, the investigation of concentrated electrolytes using so-called “NMR-exotic” nuclei ($^{35}$Cl, $^{79/81}$Br, $^{127}$I, and $^{17}$O), as well as the characterization of micrometre-scale porosity in membrane materials.

\section{STRAFI Technique}
\subsection{Experimental setup}

\begin{figure}[hbtp]
    \centering
    \includegraphics[width=\linewidth]{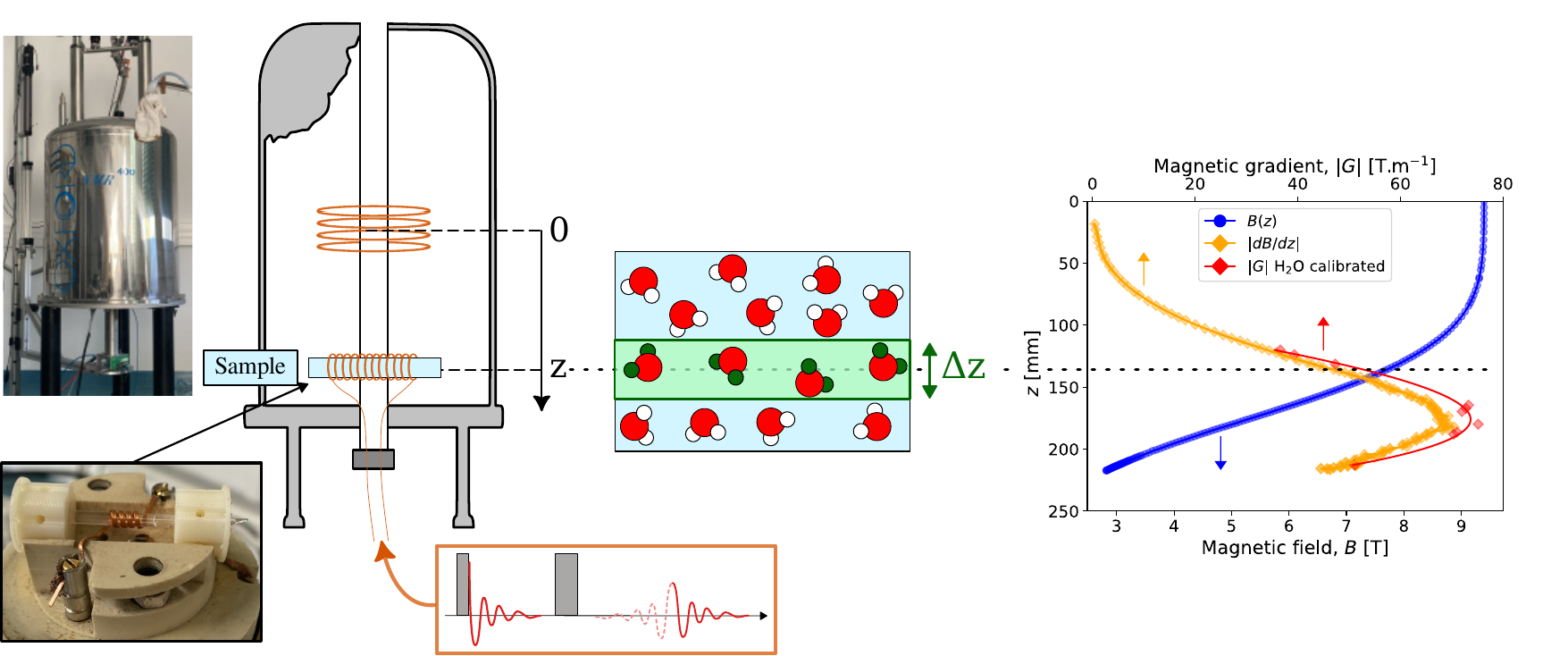}
    \caption{Schematic drawing of the set-up. The sample is placed in the leaking field of the coil. The position $z$ determines the magnetic field $B$ and the magnetic gradient $|G|$ experienced by the sample. A pulse sequence generated by an horizontal coil around the sample excites spins in a region of thickness $\Delta z$ of the order of $10$ µm up to $1$ mm.}
    \label{fig:fieldmap}
\end{figure}

Contrary to the widely used Pulse-Field Gradient (PFG) technique, STRAy Field Imagery (STRAFI) is conducted in a permanent, intense magnetic gradient. This permanent gradient is conveniently obtained by placing the sample in the stray field of the magnet. The position $z$ of the sample is controlled by a stepper motor, with a $50$ µm resolution. The magnetic field map $B(z)$, obtained by measuring the resonant frequency of $^{7}$Li as a function of the position $z$ from the coil is shown in Fig.~\ref{fig:fieldmap} (blue curve). From these data, the magnetic gradient $|dB/dZ| = |G(z)|$ is computed (orange curve). The gradient in the stray field ranges from $1$ to $70$ T/m depending on the position from the center of the  coil. The gradient experienced by the sample at a position $z$ may differ from the computed gradient due to $(x,y)$ components of the gradient. Thus, the real magnetic gradient at a given position $z$ has to be calibrated with a sample for which the diffusion coefficient is known and independent of the diffusion time. \\

The sample is positioned perpendicular to the $z$ axis. Pulse sequences are generated by a custom-built coil designed to excite the sample over a specified frequency range. The highest accessible radiofrequency is inversely proportional to both the number of turns $N$ and the coil radius $R_{\text{coil}}$, allowing the coil to be tuned to the desired frequency range. \\

\subsection{Gradient determination}

\begin{figure}[hbtp]
    \centering
    \includegraphics[width=0.5\linewidth]{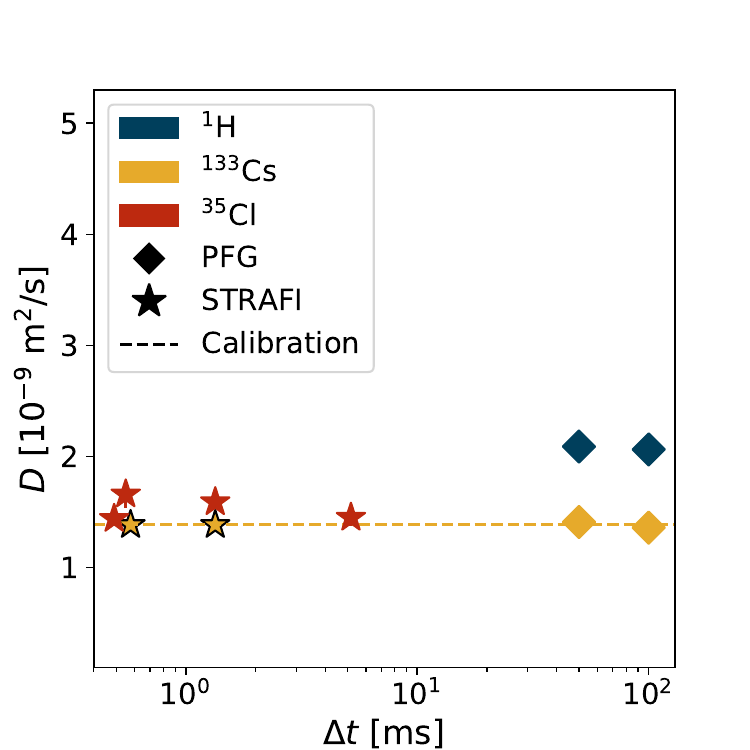}
    \caption{Calibration of the STRAFI magnetic gradient $G$ at a given position $z$ directly in the sample of interest CsCl/H$_{2}$O, using $^{133}$Cs as the probe. Diffusion of $^{133}$Cs is measured over long times $\Delta t$ using PFG-NMR experiments. This value is used to fit the magnetic gradient $G$ from STRAFI experiments. With this calibrated value, the diffusion coefficient of $^{35}$Cl in the same sample is obtained using STRAFI-NMR. $^{1}$H could also be used as a probe.}
    \label{fig:calib_strafi}
\end{figure}

A precise determination of the permanent magnetic gradient $G$ at the measurement position $z$ is essential. One approach is to use a calibration sample with a known diffusion coefficient, as illustrated in Figure~\ref{fig:fieldmap}, where water was used at different positions. A more convenient method, when feasible, is to use a nucleus within the sample of interest whose dynamics can be independently measured by PFG-NMR. The diffusion coefficient obtained from PFG-NMR is then used to fit the gradient $G$ in the STRAFI experiment. An example of this procedure is shown in Figure \ref{fig:calib_strafi}, where the sample is CsCl/H$_{2}$O and $^{133}$Cs serves as the probe nucleus. Using the precisely determined gradient, the diffusion of $^{35}$Cl in this system—which cannot be measured by PFG-NMR—can be obtained via STRAFI-NMR. In that as, $^{1}$ H could also serve as a probe in this system, since its diffusion can be independently measured by PFG-NMR as well.

\subsection{Pulse sequences}

\begin{figure}[hbtp]
    \centering
    \includegraphics[width=\linewidth]{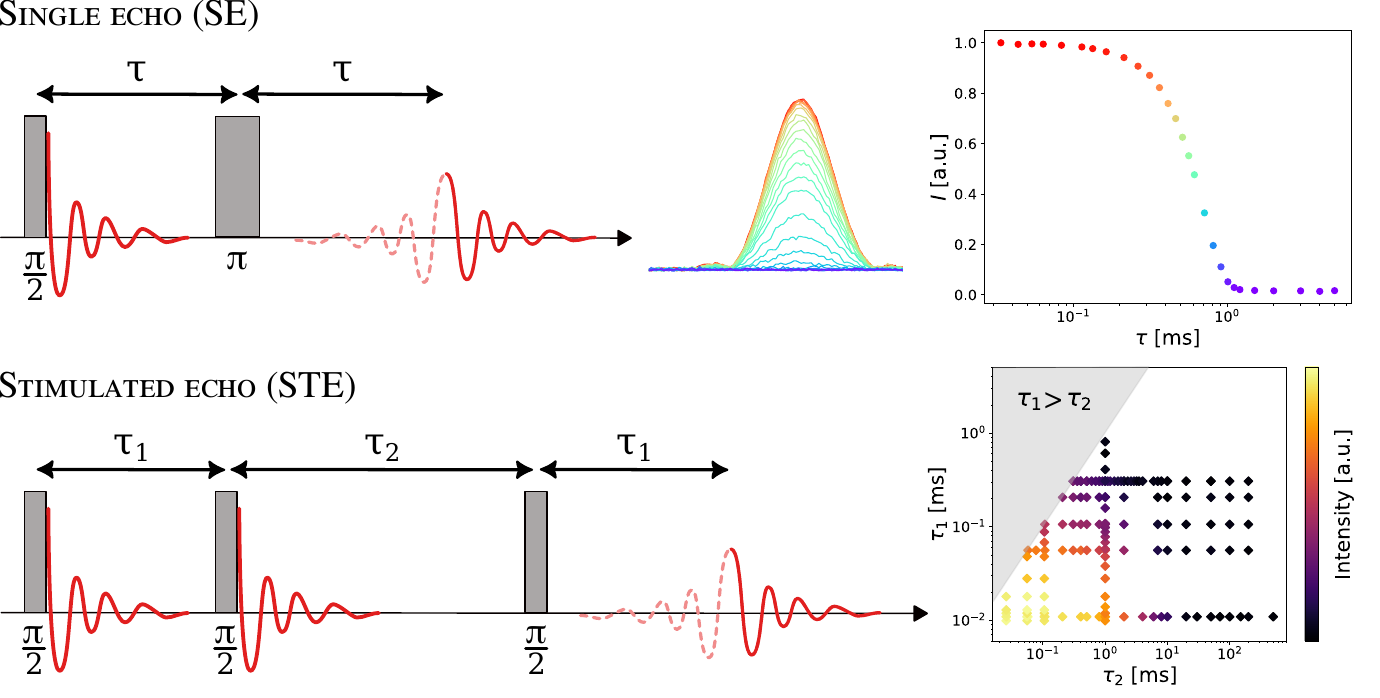}
    \caption{Single Echo (SE) (top) and Stimulated Echo (STE) (bottom) sequences. For each of them, an example of the echo intensity decrease when varying $\tau$ (top, SE on $^{17}$O in water at $298$ K) or $\tau_{1}$ and $\tau_{2}$ (bottom, STE on $^{23}$Na in a 2-methoxyethyl ether with NaTFSI salts at $333$ K) are shown on the right. The decrease of the echo intensity as a function of $\tau$ is shown for the SE sequence on the top middle figure, where colours correspond to $\tau$ values shown in the top right graph.}
    \label{fig:pulseseq}
\end{figure}

A pulse of duration $\Delta \tau_{\text{pulse}}$ excites spins within the frequency range $\Delta \nu$, corresponding — under the applied gradient $G$ — to spins located in a slice of width $\Delta z$, given by:

\begin{equation}
    \Delta z = \frac{4 \pi}{\gamma |G| \Delta \tau_{\text{pulse}}}
\end{equation}

\noindent with $\gamma$ the gyromagnetic ratio of the targeted nucleus. For a $1$ \textmu s pulse in a $60$ T/m magnetic gradient, $^{1}$H nuclei within a thickness of $\Delta z \simeq 800$ µm are excited. Single Echo (SE) and Stimulated Echo (STE) sequences can be used. The integral of the echo intensity is plotted versus the waiting time between pulses in the sequence. The decrease of the intensity is due to both spin relaxation ($T_{1}$ or $T_{2}$) and diffusion of nuclei out of the probed volume.

\section{Single Echo sequence}

For the SE sequence, the intensity decrease is given by:

\begin{equation}
    I(\tau) = I_{0} \exp {\left( \frac{-2\tau}{T_{2}} \right)} \exp {\left( -\frac{2}{3}DG^{2}\gamma^{2}\tau^{3} \right) }
\label{eq:intensitySE}
\end{equation}

\noindent with $G$ the magnetic gradient experienced by the sample and $D$ the diffusion coefficient of the nuclei \cite{hahn_spin_1950}. The time by which the $T_{2}$ relaxation term has decreased by half is given by $\tau_{T_{2}} = \frac{\ln{2}}{2} T_{2}$. Similarly, the diffusion term is decreased by half after a time $\tau_{D}^{SE}$ given by:

\begin{equation}
    \tau_{D}^{SE} = \left( \frac{3 \ln{2}}{2} \frac{1}{DG^{2}\gamma^{2}} \right)^{1/3}
\label{eq:tauDSE}
\end{equation}

In order to precisely fit the diffusion coefficient $D$, the decrease in the echo intensity due to diffusion must occur before the decrease due to the $T_{2}$ relaxation, $\tau_{D}^{SE} \lesssim \tau_{T_{2}}$. This leads to:

\begin{equation}
    T_{2} \gtrsim \left( \frac{12}{(\ln{2})^{2}} \frac{1}{DG^{2}\gamma^{2}} \right)^{1/3} = T_{2,min}^{SE}
\label{eq:T2minSE}
\end{equation}

\begin{figure}[hbtp]
    \centering
    \includegraphics[width=0.99\linewidth]{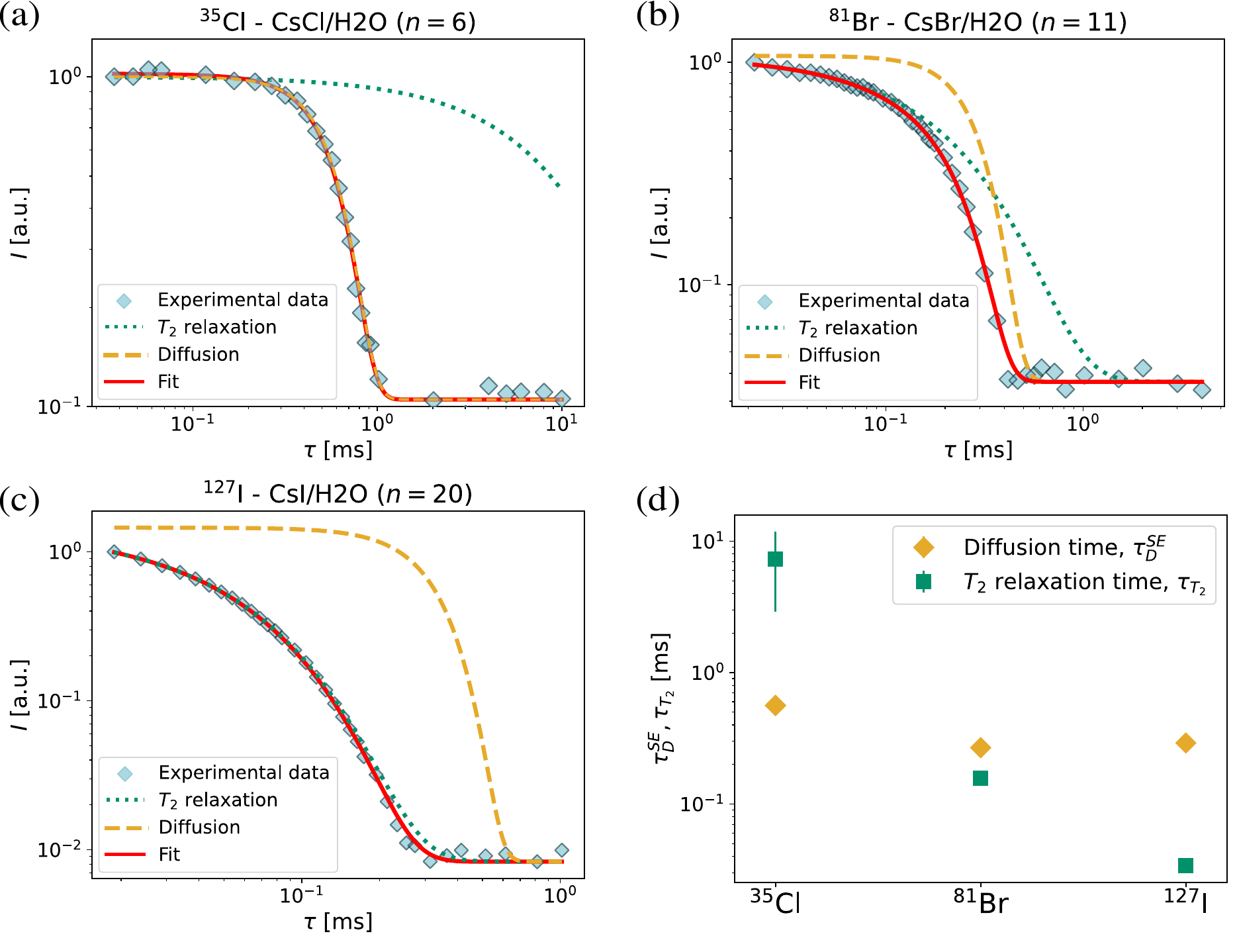}
    \caption{Fit of echo intensity decrease for SE sequences on CsX/H$_{2}$O salts at saturation concentration, $n$ being the number of water molecules per CsX at this concentration. Experimental data (blue diamonds) are fitted with Eq. \ref{eq:intensitySE}, with $T_{2}$ and $D$ as fitting parameters. The obtained values are used to compute $\tau_{D}^{SE}$ and $\tau_{T_{2}}$, which are shown (d) for the three studied nuclei $^{35}$Cl, $^{81}$Br and $^{127}$I.}
    \label{fig:fitT2vsTdiff}
\end{figure}

An example of SE sequences on three nuclei ($^{35}$Cl, $^{81}$Br and $^{127}I$) in CsX/H$_{2}$O salts at saturation concentration is shown in Fig. \ref{fig:fitT2vsTdiff}. Data are fitted by Eq. \ref{eq:intensitySE}, with $T_{2}$ and $D$ as fitting parameters. The bottom right graph shows $\tau_{D}^{SE}$ (yellow) and $\tau_{T_{2}}$ (green) for the three different nuclei. $^{35}$Cl is the favourable case for which the intensity decrease is dominated by diffusion while $T_{2}$ relaxation happens on a much slower timescale, and thus, the diffusion coefficient $D$ is well determined. $^{81}$Br is the intermediate case for which both processes occur on the same timescale and can be fitted simultaneously with a good precision. $^{127}$I is the limiting case for which most of the echo intensity decrease is due to $T_{2}$ relaxation and thus, the diffusion coefficient is poorly measured.



\section{Stimulated Echo sequence}

By combining two variable times $\tau_{1}$ and $\tau_{2}$, the Stimulated Echo (STE) sequence allows for another degree of freedom to study the dynamics of nuclei. In an STE experiment, the echo intensity varies according to
\begin{equation}
    I(\tau_{1},\tau_{2}) = \frac{I_{0}}{2} \exp {\left( \frac{-2\tau_{1}}{T_{2}} \right)} \exp {\left( \frac{-\tau_{2}}{T_{1}} \right)} \exp {\left( -DG^{2}\gamma^{2}\tau_{1}^{2}(\tau_{2}+\frac{2}{3}\tau_{1}) \right) }
\label{eq:intensitySTE}
\end{equation}
with the constraint $\tau_{1} < \tau_{2}$ \cite{hahn_spin_1950}.  
Both approaches — varying $\tau_{1}$ while keeping $\tau_{2}$ fixed, or varying $\tau_{2}$ at constant $\tau_{1}$ — are in principle possible. However, when $\tau_{1}$ is held constant, the $T_{1}$ relaxation and diffusion contributions have the same dependence on $\tau_{2}$: $\ln I \sim -\tau_{2} \left( \frac{1}{T_{1}} + D G^{2} \gamma^{2} \tau_{1}^{2} \right)$. As a result, extracting $D$ requires an accurate knowledge of $T_{1}$: any uncertainty in $T_{1}$ directly propagates into the estimation of $D$.  For this reason, it is generally safer to vary $\tau_{1}$ while keeping $\tau_{2}$ fixed, or alternatively to combine both types of experiments and perform a global fit to all datasets.\\

The diffusion time $\tau_{D}^{\mathrm{STE}}$ — defined as the time at which the diffusive attenuation term has decreased to half its initial value — is given in the Supplementary Material for an STE sequence with fixed $\tau_{2}$ and variable $\tau_{1}$. Since $\tau_{D}^{\mathrm{STE}}$ decreases as $\tau_{2}$ increases, varying $\tau_{2}$ makes it possible (1) to determine the diffusion coefficient over a wide range of diffusion times and (2) to access a regime in which diffusion dominates the $T_{2}$ relaxation, even for nuclei with very short $T_{2}$. \\

Because $\tau_{1}$ must remain smaller than $\tau_{2}$, the choice of $\tau_{2}$ sets the upper limit of the observable signal decay. To detect the intensity decrease induced by diffusion, $\tau_{1}$ must at least reach $\tau_{D}^{\mathrm{STE}}$. Therefore, one must satisfy the additional condition $\tau_{2} > \tau_{D}^{STE}$. \\

\begin{figure}
    \centering
    \includegraphics[width=0.45\linewidth]{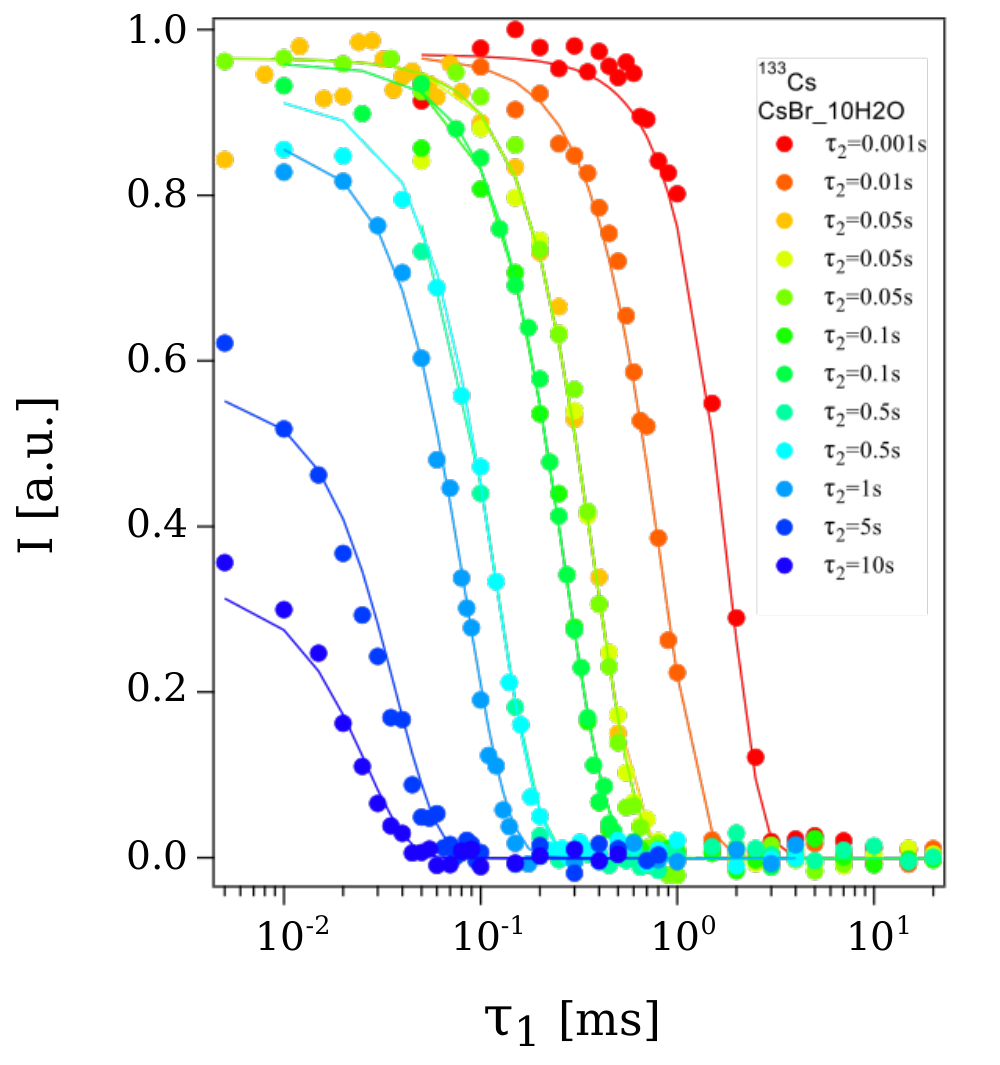}
    \caption{Simultaneous fit of STE sequences $I(\tau_{1})$ for various fixed $\tau_{2}$ values, for $^{133}$Cs in CsBr/H$_{2}$O, using a genetic algorithm. Each colour corresponds to $I(\tau_{1})$ for a given fixed $\tau_{2}$ value.}
    \label{fig:cesium2Dfit}
\end{figure}

Simultaneous fit of $I(\tau_{1},\tau_{2})$ data allows the precise determination of both the diffusion coefficient $D$ and the relaxation times $T_{1}$ and/or $T_{2}$ if they are not too long compared to the range of explored diffusion times. An example of such a fit is given in Figure \ref{fig:cesium2Dfit}, where many $I(\tau_{1})$ are obtained from STE sequences at various fixed $\tau_{2}$ values, from $\tau_{2} = 1$ ms to $1$ s. A simultaneous fit of all the sequences provides a precise measurement of both $D = 3.5  \; 10^{-9}$ m$^{2}$s$^{-1}$ and $T_{2} = 149$ ms, whereas analysing a single $I(\tau_{1})$ sequence yields only the diffusion coefficient. Depending on the complexity of the fit, standard methods such as non-linear least squares can be used, while more challenging problems may require a genetic algorithm \cite{goldberggenetic,mitchellgenetic} or other global optimization techniques. It should be noted that, because the simultaneous analysis fits data over a wide range of diffusion times using a single diffusion coefficient, it assumes Brownian diffusion---that is, a diffusion coefficient independent of the diffusion time---which is not valid for confined liquids, for example.

\section{Applications}

\subsection{Dynamics of exotic nuclei}

\begin{figure}[hbtp]
    \centering
    \includegraphics[width=\linewidth]{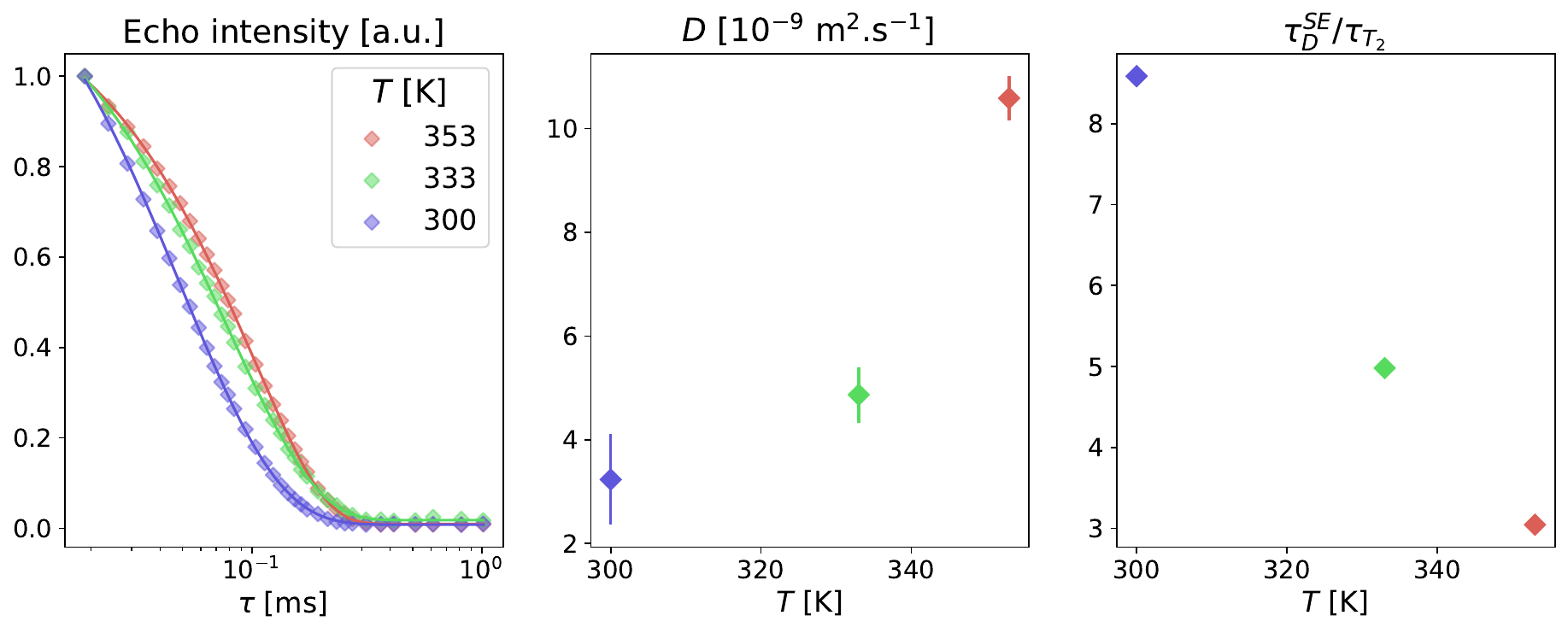}
    \caption{SE measurement of $^{127}$I in CsI/H$_{2}$O as three different temperatures. Left: Echo intensity fitted by Eq. \ref{eq:intensitySE}. Middle: Diffusion coefficient obtained from fits. Right: Ratio diffusion time over $T_{2}$ relaxation time $\tau_{D}^{SE}/\tau_{T_{2}}$ as a function of temperature.}
    \label{fig:IvsT}
\end{figure}

Because STRAFI can probe diffusion on timescales much shorter than those accessible with PFG methods, it is particularly well suited for studying the dynamics of exotic nuclei with very short $T_{2}$ relaxation times. By adjusting the gradient $G$ and selecting an appropriate pulse sequence (SE or STE with different values of $\tau_{2}$), one can reach a regime in which the diffusion time $\tau_{D}$ becomes comparable to, or even shorter than, the $T_{2}$ relaxation time $\tau_{T_{2}}$, even for nuclei exhibiting very small $T_{2}$ values. Figure~\ref{fig:IvsT} shows an example for $^{127}$I in CsI/H$_{2}$O. At $300$~K, the ratio $\tau_{D}^{SE}/\tau_{T_{2}}$ is close to $10$, which represents the upper limit at which the diffusion coefficient can still be reliably measured. Upon heating, this ratio decreases toward unity, enabling a more precise extraction of the diffusion coefficient from the fits. Consequently, the diffusion coefficient at $300$~K can be obtained by extrapolation from measurements performed at higher temperatures. \\

\subsection{Multimodal dynamics}

The wide range of diffusion times that can be probed by STRAFI experiments allows to distinguish nuclei that display multimodal dynamics. For example, for bimodal dynamics, labelling $a$ the slow population and $b$ the fast one, the decrease of the echo intensity can be expressed as:

\begin{equation}
    I = r  \tilde{I_{a}}(D_{a}, T_{1,a},T_{2,a}) + (1-r) \tilde{I_{b}}(D_{b}, T_{1,b},T_{2,b})
    \label{eq:bimodaldynamics}
\end{equation}

\noindent where $r$ is the fraction of $a$ nuclei, and $\tilde{I_{a}}$ and $\tilde{I_{b}}$ are the echo intensity decays associated with populations $a$ and $b$, respectively. These decays are given by Eq. \ref{eq:intensitySTE}, with parameters $D_{a}, T_{1,a},T_{2,a}$ for population $a$ and $D_{b}, T_{1,b},T_{2,b}$ for population $b$.\\

\begin{figure}[hbtp]
    \centering
    \includegraphics[width=0.5\linewidth]{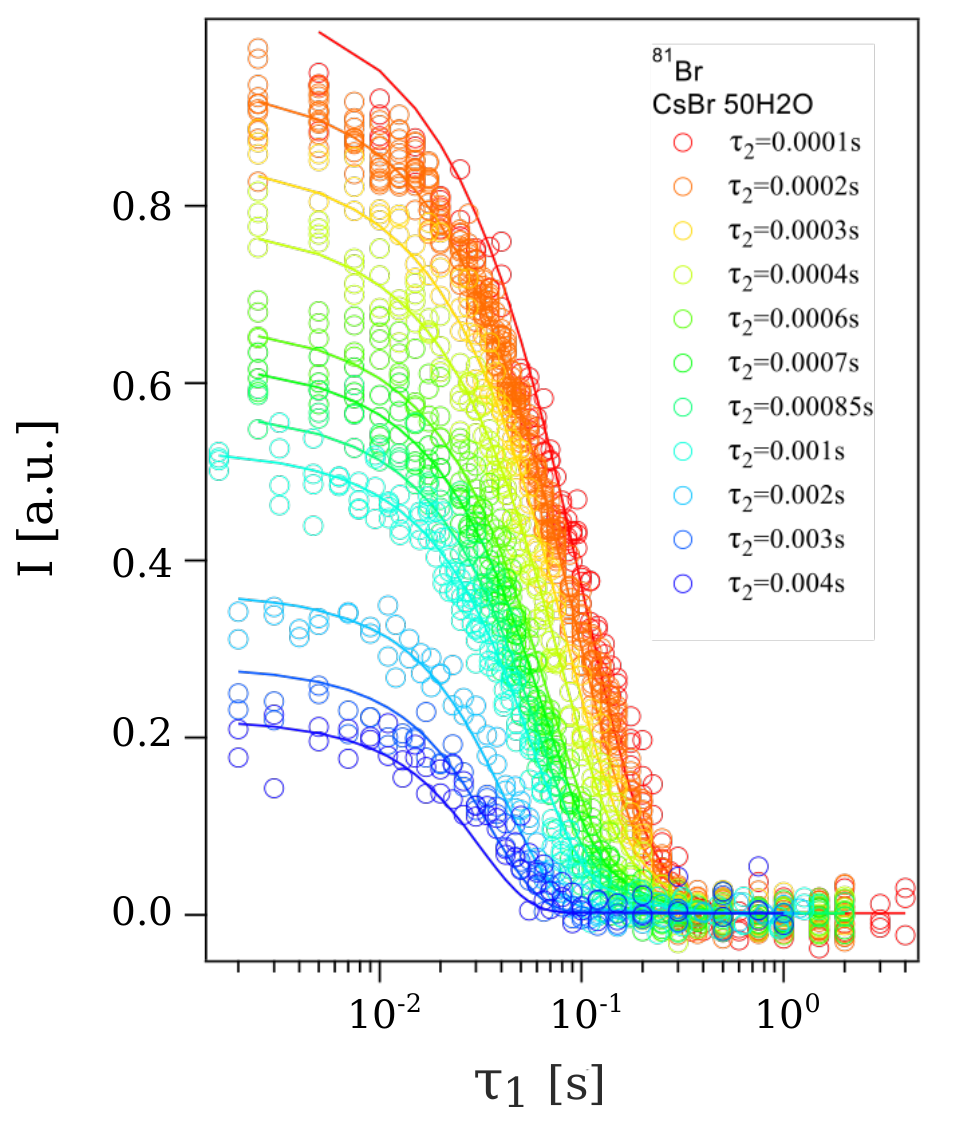}
    \caption{Bimodal dynamics of $^{81}$Br in CsBr/H$_{2}$O. Each colour corresponds to a $I(\tau_{1})$ sequence with a fixed $\tau_{2}$ value (from $100$ \textmu s (red) to $4$ ms (blue)). All these sequences are simultaneously fitted by a bimodal dynamics with Eq. \ref{eq:bimodaldynamics}, using a genetic algorithm.}
    \label{fig:bromide81_twopopulations}
\end{figure}

An example of such a situation is given in Figure \ref{fig:bromide81_twopopulations}. The dynamics of $^{81}$Br in CsBr/H$_{2}$O cannot be fitted with a monomodal dynamics. Instead, Eq. \ref{eq:bimodaldynamics} is used to simultaneously fit all the STE data. This yields a slow populations with a diffusion coefficient $D = 8.9 \; 10^{-11}$ m$^{2}$.s$^{-1}$ and a $T_{1} = 477$ µs ; and a fast population with $D = 8.3 \; 10^{-9}$ m$^{2}$.s$^{-1}$ and $T_{1} = 1.952$ ms (both population share the same $T_{2}$ value). \\

This can be extended to multimodal dynamics. The extraction of each parameter is governed by the same criterion described above: the typical diffusion and relaxation times must lie within the explored timescales in order to reliably determine the corresponding parameter.

\subsection{Diffusion in confined liquids}

The ability of STRAFI to access diffusion times below the millisecond range is not only advantageous for studying nuclei with short $T_{2}$ values, but also for probing the diffusion coefficient as a function of the diffusion time over a broad range of timescales. This is particularly relevant for confined liquids, whose diffusion behaviour depends on the probed timescale \cite{mitra1992,mitra1993,Casanova2011}. Over short times, the liquid does not yet explore the whole volume of the pores, and the diffusion is Brownian (independent on diffusion time). In this regime, the diffusion coefficient is either comparable to the bulk diffusion coefficient, or slightly different if the confinement affects other physical properties (dielectric constant, nanostructuration...). As the diffusion time increases, molecules travel longer distances. When the average molecular displacement $L$ reaches the typical pore size $d$, the measured (apparent) diffusion coefficient is lowered. The transition between these two regimes occurs at a characteristic time $\Delta t_{\text{cross}} = d^{2} / D_{\text{Brownian}}$, where $D_{\text{Brownian}}$ is the Brownian diffusion coefficient measured at short times. Analyzing how the apparent diffusion coefficient $D$ depends on the diffusion time $\Delta t$ provides information on the surface-to-pore volume ratio and the strength of molecular adsorption within the pores. This is particularly relevant for probing the dynamics of liquids confined within membranes, where transport depends sensitively on both the confinement dimensions and the interactions between the liquid and the pore surfaces. \\

\begin{figure}[hbtp]
    \centering
    \includegraphics[width=0.9\linewidth]{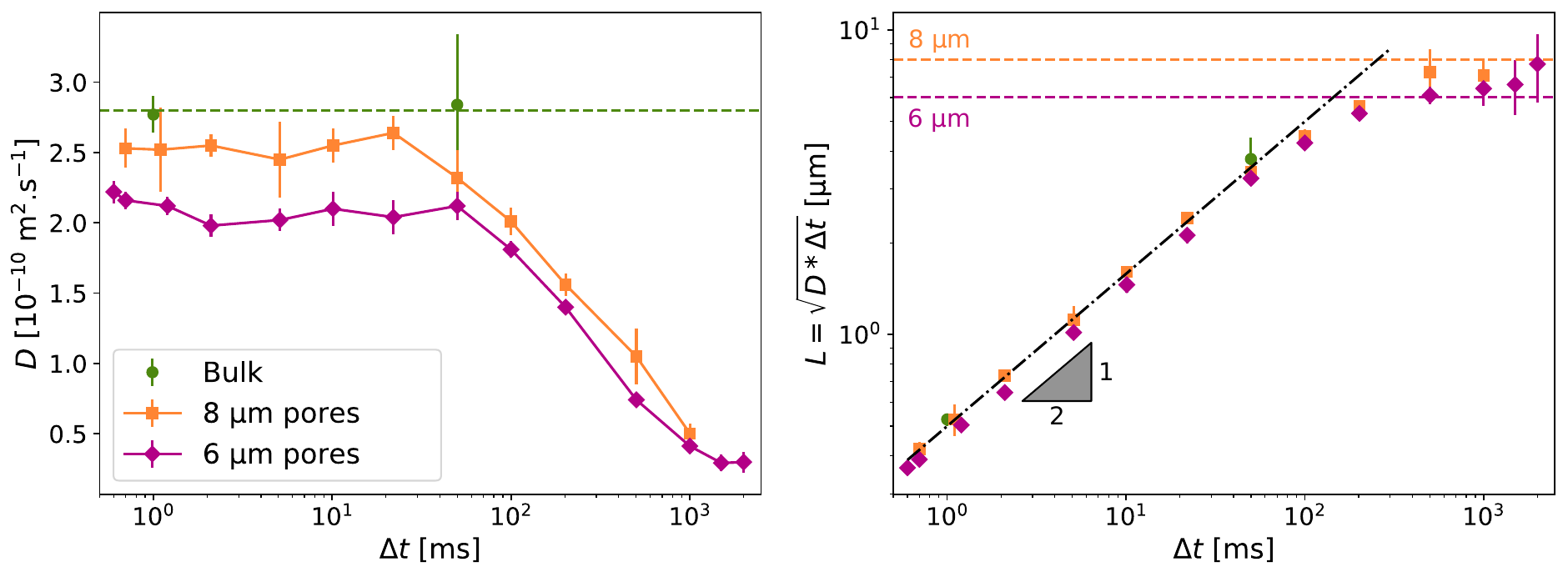}
    \caption{Diffusion coefficient of TEGdiOMe in paper filter of various porosities.}
    \label{fig:effectconfinement}
\end{figure}

An example of such a measurement is shown in Figure \ref{fig:effectconfinement}, illustrating the diffusion of TEGdiOMe (triethylene glycol dimethyl ether) in Whatman paper filters of varying porosities. At short diffusion times $\Delta t$, the molecules exhibit Brownian motion $D_{\text{Brownian}}$, corresponding to a slope $1/2$ in the log-log plot of the diffusive length $L$ with respect to $\Delta t$. The value of $D_{\text{Brownian}}$ shows slight differences between the bulk and confined liquids, which may result from strong adsorption of the liquid onto the pore surfaces, whose chemistry can vary between different paper filters. At longer times, the apparent diffusion coefficient decreases, reflecting a plateau in the average molecular displacement $L$ that corresponds to the characteristic pore size.

\section*{Conclusion}

In this work, we demonstrate that STRAFI-NMR is a powerful technique for probing the dynamics of liquids over a broad range of timescales ($50$ \textmu s to $10$ s). It is therefore well suited for measuring the diffusion coefficients of rapidly relaxing species such as $^{23}$Na and $^{127}$I. Moreover, the ability of STRAFI-NMR to define the diffusion time over a broad range of timescales is particularly advantageous for investigating confined liquids, whose dynamics depends on the probed lengthscale, and thus, on the probed diffusion time. In particular, this time dependence enables the quantification of the material's porosity and allows one to distinguish the dynamics occurring within confined regions from those occurring over much larger length scales. Finally, by combining Single Echo, Stimulated Echo sequences or more complex sequences \cite{hurlimann_diffusion_2001}, STRAFI-NMR offers multiple experimental parameters that enable accurate determination of diffusion coefficients as well as $T_{1}$ and $T_{2}$ relaxation times. \\

Looking forward, this versatility opens many opportunities for the study of complex systems. STRAFI-NMR can be applied to liquids confined in microporous materials, membranes, or biological matrices. It further enables the investigation of quadrupolar nuclei, providing unique insights into multicomponent electrolyte solutions, ionic liquids, and polymeric systems. By bridging diffusion timescales from microseconds to seconds, STRAFI-NMR offers a multiscale perspective that is difficult to achieve with conventional PFG-NMR, making it a valuable tool for both fundamental studies and technological applications in materials science, energy storage, and soft matter research.

\section*{Acknowledgments}

The authors gratefully acknowledge the support of Labex PALM and CNRS. They also thank the Elinstru team for their contributions to the experimental development of the STRAFI method.

\bibliographystyle{siam}  
\bibliography{biblio}      

\end{document}